\documentclass[12pt]{article}

\usepackage{amsmath,amsfonts,amssymb}

\def\mM{\mathcal{M}}

\def\tE{\tilde{E}}

\def\hphi{\hat{\phi}}

\def\balpha{\bar{\alpha}}
\def\bbeta{\bar{\beta}}

\def\ty{\tilde{y}}

\def\mB{\mathcal{B}}
\def\bA{\mathbf{A}}

\def\tx{\tilde{x}}

\def\be{\begin{equation}}

\def\halpha{\hat{\alpha}}
\def\hbeta{\hat{\beta}}
\def\ee{\end{equation}}

\def\bea{\begin{eqnarray}}

\def\eea{\end{eqnarray}}

\def\tr{\mathrm{tr}\, }

\def\bmu{\bar{\mu}}
\def\bnu{\bar{\nu}}

\def\bI{\mathbf{I}}

\def\tr{\mathrm{Tr}}

\def\str{\mathrm{Str}}

\newcommand{\mF}{\mathcal{F}}

\newcommand{\mU}{\mathcal{U}}

\def \bA{\mathbf{A}}

\def\hnu{\hat{\nu}}
\def\hmu{\hat{\mu}}

\newcommand{\tphi}{\tilde{\phi}}

\begin{document}

	\begin{titlepage}

		\vskip 0.4 cm
		
		\begin{center}
			{\Large{ \bf
					
				D-Brane Actions in Non-Relativistic String Theory and
			T-Duality}}
			
			\vspace{1em}  Josef Kluso\v{n}$\,^1$
			\footnote{Email address:
				klu@physics.muni.cz}\\
			\vspace{1em} $^1$\textit{Department of Theoretical Physics and
				Astrophysics, Faculty
				of Science,\\
				Masaryk University, Kotl\'a\v{r}sk\'a 2, 611 37, Brno, Czech Republic}\\

			%
			%
			
			\vskip 0.8cm
			
		\end{center}

		\begin{abstract}
We study D-brane actions in non-relativistic string theory. We consider single D-brane and analyse its properties under T-duality. We also derive an action for $N$ D8-branes and also for lower dimensional D-branes through T-duality transformations. 			
		\end{abstract}
		
		\bigskip
		
	\end{titlepage}
	
	\newpage

\section{Introduction and Summary}\label{first}
Non-relativistic string theory is specific form of two dimensional quantum field theory defined on the string world-sheet \cite{Gomis:2000bd,Danielsson:2000gi}. The target space geometry of this two-dimensional field theory is stringy Newton-Cartan geometry, see also 
\cite{Gomis:2020izd,Gomis:2020fui,Bergshoeff:2019pij,
	Harmark:2019upf,Gomis:2019zyu,Kluson:2019ifd,Kluson:2018vfd,
	Bergshoeff:2018yvt}.

Stringy Newton-Cartan geometry can be physically defined as a theory where  probe of the geometrz  is two dimensional object (string). As a result target space-time naturally splits into two dimensional longitudinal sector and $8-$dimensional transverse sector where we will implicitly work with the parent string theory with critical dimension equal to ten. As is well known from the study of non-relativistic string theory \cite{Gomis:2000bd,Danielsson:2000gi} in the flat space-time longitudinal 
sector is Lorentzian while transverse sector is Euclidean.

It is very interesting to analyse open string theory in this background where now boundary conditions imposed at the end of string world-sheet play an important role. Very careful analysis of open string theory in non-relativistic background was performed recently in \cite{Gomis:2020izd,Gomis:2020fui}. It was shown there that imposing Dirichlet boundary conditions in longitudinal sector leads to non-relativistic open string theory on  D8-brane. It was shown in
\cite{Gomis:2020fui}
 that 
in the flat space-time low energy effective action of open strings on $N$-non-relativistic D8-branes is described by Galilean invariant $U(N)-$Yang-Mills theory. T-duality of non-relativistic open string theory was further analysed in 
\cite{Gomis:2020izd} firstly in the flat space-time and then it was studied in 
general space-time using sigma model description or using Dirac-Born-Infeld (DBI) description. The goal of this paper is to extend the analysis of T-duality of non-relativistic open string using DBI description and its non-abelian generalization.

We start our paper with the standard construction of non-relativistic Dp-brane. In more details, we consider DBI action for string in the background with light-like isometry and we presume that Dp-brane is extended along this direction. Then we perform  dimensional reduction along this direction
\footnote{We could presume that light-like direction is compact that can be defined by infinite boosting longitudinal spatial  circle.}. As a result we obtain non-relativistic D(p-1)-brane which is generalization of the construction presented in \cite{Gomis:2020izd} when some of the direction in transverse sector obey Dirichlet boundary conditions. Note that using this construction we get D(p-1)-brane that is localized at light-like direction. We further study 
T-duality along $k-$spatial directions transverse to the world-volume of D(p-1)-brane and we obtain an action for non-relativistic  D(p-1-k)-brane 
in the background that is related to the original one by  Buscher's rules  \cite{Ginsparg:1992af,Quevedo:1993vq} which are generalization of standard 
Buscher's prescription   \cite{Buscher:1987sk,Buscher:1987qj} to the case of T-duality along $k-$ directions
\footnote{For earlier review of T-duality, see \cite{Giveon:1994fu}.}. However it is important to stress that the resulting D(p-1-k)-brane action has the same form as non-relativistic D(p-1)-brane on condition when the components of time form $\tau$  along spatial directions that we dualize, are zero. Note that this is the same condition that was imposed in \cite{Bergshoeff:2018yvt}
when T-duality of closed string along transverse directions was analysed. 

As the next step we proceed to the construction of an action for $N$ non-relativistic Dp-branes. To do this we closely follow \cite{Myers:1999ps}. Explicitly, we start with the space-time filling DBI action for $N$ D9-branes in the background with like-like isometry. Performing T-duality along light-like direction we obtain action for $N$ non-relativistic D8-branes. An action for $N$ D(8-k)-branes is derived by performing T-duality along $k-$spatial dimensions. This action is crucial for the analysis of T-duality along direction transverse to the D-brane world-volume. In more details, in order to study T-duality of Dp-brane along directions transverse to its world-volume we should consider an infinite array of Dp-branes on the covering space when all world-volume fields obey quotient conditions 
\cite{Taylor:1996ik,Taylor:1997dy,Brace:1998ku}. We firstly apply this approach to the case of non-relativistic D8-brane transverse to light-like direction. We replace it with the action for infinite D8-branes on the covering space and we find that this configuration is equivalent to D9-brane in the background with light-like direction. Then we generalize this approach to the case of T-duality along light-like direction transverse to D(8-k)-brane and we again show that it is equivalent to D(9-k)-brane where this D(9-k)-brane is extended along light-like direction. 

Let us outline or results and suggest possible extension of this work. We study an effective actions for non-relativistic Dp-branes. We find their form and we analyse as their transform under T-duality transformations. We then generalize this result to the case of $N$ non-relativistic D8 and D(8-k)-branes. We argue that these non-abelian generalizations are crucial for analysis how non-relativistic D(8-k)-brane transforms under T-duality along transverse light like directions when we show that non-relativistic D(8-k)-brane maps to relativistic D(9-k)-brane in the background with light-like isometry. We mean that this is nice consistency check of the proposal of the action for $N$ non-relativistic D(8-k)-branes. 

The natural extension of this work would be to analyse properties of the Wess-Zumino term  which describes how Dp-brane couples to Ramond-Ramond forms. However in order to do this we should know non-relativistic limit of Ramond-Ramond fields which has not been found yet. On the other hand it would be natural to start
with the fact that non-relativistic D(p-1)-brane is defined using T-duality of relativistic Dp-brane along light-like directions. Then we should perform the same T-duality transformations in case of Wess-Zumino term and we could find term that expresses coupling of non-relativistic D(p-1)-brane to Ramond-Ramond form. We hope to return to this problem in future. 

This paper is organized as follows. In the next section (\ref{second}) we introduce non-relativistic D(p-1)-brane. Then in section (\ref{third}) we study its properties under T-duality. In section (\ref{fourth}) we perform its non-abelian generalization. Finally in section (\ref{fifth}) we study how non-relativistic D-branes transform under T-duality performed along light-like directions which is transverse to their world-volumes.

\section{Non-Relativistic D-Brane and T-Duality}\label{second}
In this section we review and extend construction of Dp-branes in non-relativistic background, following recent work \cite{Gomis:2020izd}.

We start with  Dp-brane action  in the relativistic background with light like isometry along $y-$ direction which means that the metric component $g_{yy}=0$ and all space-time fields do not depend on $y$. Let us now consider DBI action for Dp-brane in general background   
\begin{eqnarray}\label{DBIaction}
& &S=-T_p\int d^{p+1}\xi e^{-\phi}\sqrt{-\det \bA_{\alpha\beta}} \ , \nonumber \\ 
& &\bA_{\alpha\beta}=g_{\mu\nu}\partial_\alpha x^\mu \partial_\beta x^\nu+
b_{\mu\nu}\partial_\alpha x^\mu\partial_\beta x^\nu+\lambda F_{\alpha\beta} \ , 
\end{eqnarray}
where $\xi^\alpha,\alpha,\beta=0,1,\dots,p$ label world-volume of p-brane and where 
$x^\mu(\xi),\mu=0,1,\dots,9$ are world-volume fields that describe embedding of 
Dp-brane in the target space-time with the dilaton $\phi$, metric $g_{\mu\nu}$ and NSNS two form field $b_{\mu\nu}$. Further, $F_{\alpha\beta}=\partial_\alpha A_\beta-\partial_\beta A_\alpha$, where $A_{\alpha}$ is gauge field propagating on the world-volume of Dp-brane. Finally $\lambda=2\pi\alpha'$ and $T_p=\frac{1}{\lambda^{(p+1)/2}}$ is Dp-brane tension.

Let us presume that Dp-brane is extended along $y\equiv x^9$ direction so that we impose following static gauge 
\begin{equation}
y=\xi^p \ . 
\end{equation}
Next step is to perform dimensional reduction when we presume that all world-volume
fields do not depend on $y$. As a result we obtain following form of the matrix 
 $\bA_{\alpha\beta}$
\begin{eqnarray}\label{bAdim}
& &\bA_{yy}=0 \ , \quad 
\bA_{y\hbeta}=g_{yj}\partial_{\hbeta}x^j+b_{yj}\partial_{\hbeta}x^j-\lambda\partial_{\hbeta}A_y \ , \nonumber \\
& &\bA_{\halpha y}=\partial_{\halpha} x^ig_{iy}+\partial_{\halpha} x^i b_{iy}+
\lambda \partial_{\halpha} A_y \ , \nonumber \\
& &\bA_{\halpha\hbeta}=g_{ij}\partial_{\halpha} x^i\partial_{\hbeta}x^j+
b_{ij}\partial_{\halpha}x^i\partial_{\hbeta}x^j+\lambda F_{\halpha\hbeta} \ , \nonumber \\
\end{eqnarray}
where $\xi^{\halpha},\halpha,\hbeta=0,1,\dots,p-1$ label world-volume coordinates
on D(p-1)-brane and where $i,j=0,1,\dots,8$.

Using (\ref{bAdim}) we find that the action (\ref{DBIaction}) 
 has the form 
\begin{equation}\label{Snon}
S=-T_p\int dy\int d^{p}\xi e^{-\phi}\sqrt{-\det \left(\begin{array}{cc}
	0 & (g_{yj}+b_{yj})\partial_{\hbeta} x^j
	-\lambda\partial_{\hbeta}A_y \\
\partial_{\halpha}x^i (g_{iy}+b_{iy})+
\lambda \partial_{\halpha}A_y	& \bA_{\halpha\hbeta}\end{array}\right)} \ . 
\end{equation}
Note that the determinant in the action  (\ref{Snon}) has the form
\begin{equation}
\det \left(\begin{array}{cc}
0 & \mM_{y\hbeta} \\
\mM_{\halpha y} & 
\mM_{\halpha\hbeta} \end{array}\right)
\end{equation}
that, using properties of the determinant and also the fact that $\mM_{yy}=0$ 
is equal to 
\begin{equation}\label{detprop}
\det \left(\begin{array}{cc}
0 & \mM_{y\hbeta} \\
\mM_{\halpha y} & 
\mM_{\halpha\hbeta} \end{array}\right)=
\det \left(\begin{array}{cc}
0 & \mM_{y\hbeta} \\
\mM_{\halpha y} & 
\mM_{\halpha\hbeta}+V_{\halpha}M_{y\hbeta}+M_{\halpha y}W_{\hbeta} \\
\end{array}\right) \ , 
\end{equation}
where $V_{\halpha}$ and $W_{\hbeta}$ are arbitrary functions. 
To proceed further we define $T-$dual coordinate $\ty$ as 
\begin{equation}
\ty=\lambda  A_y
\end{equation}
and, following \cite{Gomis:2020izd} we define $\tau_{\mu}^{ \ A}, A=0,1$ as
\begin{equation}
\tau_i^{ \ 0}=Cg_{yi} \ , \quad   \tau_{\ty}^{ \ 0}=0 \ ,  \quad 
\tau_i^{ \ 1}=Cb_{yi} \ , \quad \tau_{\ty}^{ \ 1}=C
\end{equation}
so that we can write
\begin{eqnarray}
\mM_{\halpha y}=\tau_{\halpha} \ , \quad 
\mM_{y \hbeta}=\bar{\tau}_{\hbeta} \ , \quad 
\tau_{\halpha}=\tau_{\halpha}^{ \ 0}+\tau_{\halpha}^{ \ 1} \ ,  \quad 
	\bar{\tau}_{\halpha}=\tau_{\halpha}^{ \ 0}-\tau_{\halpha}^{ \ 1} \ , 
	\end{eqnarray}
	where $\tau_{\halpha}=\tau_{\mu}\partial_{\halpha}\tx^\mu=
	\tau_i
	\partial_{\halpha}x^i+\tau_{\ty} \partial_{\halpha}\ty$. Let us now define  $V_{\halpha}$ and $W_{\halpha}$ as 
\begin{eqnarray}\label{V}
& &V_{\halpha}=\partial_{\halpha}\tx^\mu V_\mu=
\partial_{\halpha}x^i(C_i^{ \ 0}+C_i^{ \ 1})+\partial_{\halpha}\ty (C_{\ty}^{ \ 0}+C_{\ty}^{ \ 1}) \ , \nonumber \\
& & W_{\halpha}=\partial_{\halpha}\tx^\mu W_{\mu}=\partial_{\halpha}x^i (C_i^{ \ 0}-C_i^{ \ 1})+\partial_{\halpha}\ty (C_{\ty}^{ \ 0}-C_{\ty}^{ \ 1}) \ .
\nonumber \\
\end{eqnarray}
Now using (\ref{V}) we obtain
\begin{eqnarray}
& &\mM_{\halpha\hbeta}+V_{\halpha}M_{y
\hbeta}+M_{\halpha y}W_{\hbeta}=\nonumber \\
& & H_{\halpha\hbeta}+\mB_{\halpha\hbeta}+\lambda F_{\halpha\hbeta} \ , \nonumber \\
\end{eqnarray}
where $H_{\halpha\hbeta}=H_{\mu\nu}\partial_{\halpha}\tx^\mu
\partial_{\hbeta} \tx^\nu$ where $H_{\mu\nu}$ is equal to  
\begin{eqnarray}
& &H_{ij}=
g_{ij}+C_i^{ \ 0}\tau_j^{ \ 0}+\tau_i^{ \ 0}C_j^{ \ 0}-C_i^{ \ 1}\tau_j^{ \ 1}
-\tau_i^{\ 1}C_j^{ \ 1} \ , \nonumber \\
& &H_{\ty\ty}=2(C_{\ty}^{ \ 0}\tau_{\ty}^{ \ 0}-C_{\ty}^{\ 1}\tau_{\ty}^{ \ 1}) \ , \nonumber \\
& & H_{\ty j}=C_{\ty}^{ \ 0}\tau_j^{ \ 0}+\tau_{\ty}^{\ 0}C_j^{ \ 0}
-C_{\ty}^{ \ 1}\tau_{j}^{ \ 1}-\tau_{\ty}^{ \ 1}C_{j}^{ \ 1} \ , \nonumber \\
& &H_{i\ty}=C_i^{ \ 0}\tau_{\ty}^{ \ 0}+\tau_i^{ \ 0}C_{\ty}^{ \ 0}-C_i^{ \ 1}\tau_{\ty}^{  \ 1}-\tau_i^{ \ 1}C_{\ty}^{\ 1} \ . \nonumber \\
\end{eqnarray}
In the same way we find that $\mB_{\mu\nu}$ is equal to
\begin{eqnarray}
& &\mB_{ij}=b_{ij}+C_i^{ \ 1}\tau_j^{ \ 0}-C_i^{ \ 0}\tau_j^{ \ 1}+
\tau_i^{ \ 1}C_j^{ \ 0}-C_i^{ \ 0}\tau_j^{ \ 1} \ , \nonumber \\
& &\mB_{i\ty}=C_i^{ \ 1}\tau_{\ty}^{ \ 0}-C_i^{ \ 0}\tau_{\ty}^{ \ 1}+
\tau_i^{ \ 1}C_{\ty}^{ \ 0}-\tau_i^{ \ 0}C_{\ty}^{ \ 1} \ , \nonumber \\
& &\mB_{\ty j}=
C^{ \ 1}_{\ty}\tau_j^{ \ 0}+\tau_{\ty}^{ \ 1}C_j^{ \ 0}
-C_{\ty}^{ \ 0}\tau_j^{ \ 1}-\tau_{\ty}^{ \ 0}C_j^{ \ 1}
\nonumber \\
\end{eqnarray}
so that the action for D(p-1)-brane has the form
\begin{eqnarray}\label{Sfinal}
S
=
-T_{p-1} \int d^p \xi e^{-\hat{\phi}}
\sqrt{-\det \left(\begin{array}{cc}
	0 & \tau_{\mu}\partial_{\hbeta}\tx^\mu \\
	\partial_{\halpha}\tx^\nu \bar{\tau}_\nu	& H_{\halpha\hbeta}+
	\mB_{\halpha\hbeta}+\lambda F_{\halpha\hbeta}\end{array}\right)} \ ,
\nonumber \\
\end{eqnarray}
where we identified $T_{p-1}$ as
\begin{equation}
T_{p-1}=T_p\int dy 
\end{equation}
and also $\hat{\phi}$ as
\begin{equation}
\hat{\phi}=\phi+\ln |C| \  
\end{equation}
which is the same result as was found in 
\cite{Gomis:2020izd}.  Finally using properties of determinant 
(\ref{detprop}) we see that there is natural redefinition of the background fields given by prescription 
\begin{equation}
\tilde{H}_{\halpha\hbeta}+\tilde{\mB}_{\halpha\hbeta}
=H_{\halpha\hbeta}+\mB_{\halpha\hbeta}+
X_{\halpha}\bar{\tau}_{\hbeta}+\tau_{\halpha}Y_{\hbeta} \  
\end{equation}
that can be written in equivalent form
\begin{eqnarray}
& &\tilde{H}_{\mu\nu}=H_{\mu\nu}-Z_\mu^{ \ A}\tau_\nu^{ \ B}\eta_{AB}-\tau_\mu^{ \ A}Z_\nu^{ \ B}\eta_{AB} \ , \nonumber \\
& &\tilde{\mB}_{\mu\nu}=\mB_{\mu\nu}-Z_\mu^{ \ A}\tau_\nu^{ \ B}\epsilon_{AB}
-\tau_\mu^{ \ A}Z_\nu^{ \ B}\epsilon_{AB} \ , 
\end{eqnarray}
where we defined $Z^{ \ A}_\mu, A=0,1$ as
\begin{equation}
X_\mu=Z_\mu^{ \ 0}+Z_\mu^{ \ 1} \ , \quad 
Y_\mu=Z_\mu^{ \ 0}-Z_\mu^{ \ 1} \ ,
\end{equation}
and where $\eta_{AB}=\mathrm{diag}(-1,1)$ and $\epsilon_{01}=-\epsilon_{10}=1$. These transformations are known as Stuckelberg transformations of the non-relativistic fields \cite{Gomis:2020fui,Gomis:2020izd,Bergshoeff:2019pij,Bergshoeff:2018yvt}.

\section{T-Duality of Non-Relativistic D(p-1)-Brane}\label{third}
In this section we study how  non-relativistic D(p-1)-brane action 
(\ref{Sfinal}) transforms under T-duality transformations. We start with situation 
when we perform T-duality along $k-$ longitudinal spatial dimensions where D(p-1)-brane
wraps them.
Then it is natural to  perform gauge fixing
\begin{equation}
\xi^m=x^m \ , m=9-k,\dots,8 \ , 
\end{equation}
where now  all world-volume fields do not depend on $\xi^m$. Instead they are functions of remaining  world-volume coordinates $\xi^{\balpha}$ where $\balpha=0,\dots,p-1-k$. Further, let us denote remaining coordinates as $x^{\bmu} \ , \bmu,\bnu=0,1,\dots,9-k$. Finally we introduce $E_{\mu\nu}=H_{\mu\nu}+\mB_{\mu\nu}$. 
In this case the action (\ref{Sfinal}) has the form
\begin{eqnarray}\label{STdualhelp}
& &S=-T_{(p-1)}\int d^{k}\xi \times \nonumber \\
& &\times \int d^{p-k}\xi e^{-\hat{\phi}}
\sqrt{-\det \left(\begin{array}{ccc}
0 & \tau_{\bmu}\partial_{\bbeta}x^{\bmu} & \tau_n \\
\partial_{\balpha}\tx^{\bmu}\bar{\tau}_{\bmu} & E_{\balpha\bbeta}+\lambda F_{\balpha\bbeta} & E_{\balpha n}+\lambda 
\partial_{\balpha} A_n \\
\bar{\tau}_m & E_{m\bbeta}-\lambda 
\partial_{\bbeta}A_m & E_{mn} \\ \end{array}\right)}=
\nonumber \\
& &=-T_{(p-1-k)}\int d^{p-k}\xi e^{-\hat{\phi}}\sqrt{\det E_{mn}}\times \nonumber \\
& &\sqrt{-\det \left(\begin{array}{cc}
	-\tau_n\tE^{nm}\bar{\tau}_m & \tau_{\bmu}\partial_{\bbeta}x^{\bmu} 
	-\tau_m \tE^{mn}(E_{n\bbeta}-\lambda\partial_{\bbeta}A_n)  
		 \\
	\partial_{\balpha}\tx^{\bmu}\bar{\tau}_{\bmu}-
	(E_{\balpha n}+\lambda \partial_{\balpha}A_n)\tE^{nk}\bar{\tau}_k	
	 & E_{\balpha\bbeta}+\lambda F_{\balpha\bbeta}-
(E_{\balpha n}+\lambda \partial_{\balpha}A_n)\tE^{nm}
(E_{m\bbeta}-\lambda \partial_{\bbeta}A_m)	 
	  \\ \end{array}\right)} \ , \nonumber \\
\end{eqnarray}
where $\tE^{mn}$ is matrix inverse to $H_{mn}+\mB_{mn}$. We see that
the form of the action for D(p-1-k)-brane does not have the form 
of non-relativistic action as was introduced in the second question due to the presence of the expression $-\tau_n\tE^{nm}\bar{\tau}_m$. However when we derived the
action (\ref{STdualhelp})  we presumed
that D(p-1)-brane wraps $x^k$ directions which are pure spatial. Then it is natural to presume that 
\begin{equation}\label{taucon}
\tau_m=\bar{\tau}_m=0 \ . 
\end{equation}
Note also that these conditions were imposed when T-duality along pure spatial dimension was analysed in 
\cite{Bergshoeff:2018yvt}.

With the help of  conditions (\ref{taucon}) we obtain that
the action for non-relativistic D(p-1-k)-brane has the form 
\begin{equation}
S=-T_{p-1-k}\int d^{p-k}\xi e^{-\tphi}
\sqrt{\det \left(\begin{array}{cc}
0 & \tau_{\bmu}\partial_{\bbeta}x^{\bmu} 
	\\
	\partial_{\balpha}\tx^{\bmu}\bar{\tau}_{\bmu}	
	& P[\tE_{\balpha\bbeta}]+\lambda F_{\balpha\bbeta}
		  \\
	\end{array}\right)} \ , 
\end{equation}
where $\tphi$ is transformed dilaton field defined by equation
\begin{equation}\label{Tdualphi}
e^{-\tphi}=e^{-\hat{\phi}}\sqrt{\det E_{mn}} \ , 
\end{equation}
and where $P[\tE_{\balpha\bbeta}]$ is pull-back of the T-dual metric to the world-volume of D(p-1-k)-brane defined as
\begin{equation}
P[\tE_{\balpha\bbeta}]=
\partial_{\balpha}x^{\bmu}\tE_{\bmu\bnu}\partial_{\bbeta}x^{\bnu}+
\partial_{\balpha}x^{\bmu}\tE_{\bmu}^{ \ m}\partial_{\bbeta}\tx_m+
\partial_{\balpha}\tx_m \tE^m_{ \ \bnu}\partial_{\bbeta}x^{\bnu}+
\partial_{\balpha}\tx_m\tE^{mn}\partial_{\bbeta}\tx_n \ , 
\end{equation}
where T-dual coordinates $\tx_m$ are defined as 
\begin{equation}
\tx_m=\lambda A_m \ . 
\end{equation}
Finally components of T-dual metric have the form 
\begin{eqnarray}\label{Tdualemtr}
\tE_{\bmu\bnu}=E_{\bmu\bnu}-E_{\bmu m}\tE^{mn}E_{n\bnu} \ ,  
\quad 
\tE_{\bmu}^{ \ m}=E_{\bmu m}\tE^{mn} \ , \quad 
\tE^n_{ \ \bnu}=-\tE^{nm}E_{m\bnu} \ . 
\end{eqnarray}
Note that (\ref{Tdualemtr}) are standard T-duality rules as are known in relativistic
string theories \cite{Ginsparg:1992af,Quevedo:1993vq}.

As the natural step we should analyse T-duality properties of non-relativistic
D(p-1)-brane when we dualize along directions transverse to the world-volume
of D(p-1)-brane. However as was shown in 
\cite{Taylor:1996ik,Taylor:1997dy} this can be done when we consider infinite number of D(p-1)-branes on the covering space whose world-volume fields obey appropriate quotient conditions. In order to do this we should firstly find
an action for $N-$ Dp-branes in non-relativistic background.

\section{Non-Abelian Generalization}\label{fourth}
In order to find non-abelian action for $N$ Dp-branes in non-relativistic background we will follow 
\cite{Myers:1999ps} when we start with the non-abelian action for $N-$space-time
filling D9-branes. This action has the form
\begin{equation}
S=-\mathrm{sTr}T_9
\int d^{10}\xi e^{-\phi}\sqrt{-\det (E_{\mu\nu}+\lambda F_{\mu\nu})} \ , 
\end{equation}
where 
\begin{equation}
F_{\mu\nu}=\partial_\mu A_\nu-\partial_\nu A_\mu+i[A_\mu,A_\nu] \ , 
\end{equation}
where $A_\mu$ are $N\times N$ Hermitian matrices and where $\mathrm{sTr}$ means
symmetrized trace.
Now as in the second section (\ref{second}) we presume that the background has light like isometry and we label this dimension by  $\xi^9\equiv y$. We further presume that all world-volume fields do not
depend on them. Then we get
\begin{equation}
\lambda F_{\hmu y}=D_{\hmu} \Phi_{\ty}=-\lambda F_{y\hmu} \ , 
\end{equation}
where we introduced $\Phi_{\ty}$ as $N\times N$ matrix through the formula
\begin{equation}
\Phi_{\ty}=\lambda A_y \ ,
\end{equation}
and where $\hmu=0,1,\dots,8$.
Note that $\Phi_{\ty}$ describes embedding of D8-branes in transverse $\ty-$ directions. Using this notation we get
\begin{equation}
\det (E_{\mu\nu
}+\lambda F_{\mu\nu})=
\det\left(\begin{array}{cc}
0 & E_{ y\hnu}-D_{\hnu} \Phi_{\ty} \\
E_{\hmu y}+D_{\hmu}\Phi_{\ty} & E_{\hmu\hnu}+\lambda F_{\hmu\hnu}
\end{array}\right) \ . 
\end{equation}
Then  performing the same manipulation as in the second section (\ref{second}) 
we obtain an action for $N$ D8-branes in non-relativistic background
in the form 
\begin{equation}\label{ND8}
S=-T_8 \int d^9\xi \str e^{-\hat{\phi}}
\sqrt{-\det\left(\begin{array}{cc}
	0 & \tau_{\hat{\nu}}+\tau_{\ty}D_{\hnu}\Phi_{\ty} \\
\bar{\tau}_{\hmu}+D_{\hmu} \Phi_{\ty} \bar{\tau}_{\ty} & 
H_{\hmu\hnu}+\mB_{\hmu\hnu}+\lambda F_{\hmu\hnu} \end{array}\right)} \ .
\end{equation}
Let us now perform T-duality along $k-$directions which means
that all world-volume fields do not depend on $\xi^m, m=(9-k),\dots,8$. 
Then $\bmu,\bnu=0,1,\dots,8-k$ are world-volume coordinates.
As a result we obtain 
\begin{eqnarray}
F_{\bmu n}=D_{\bmu}\Phi_n \ , \quad 
F_{m\bnu}=-D_{\bnu}\Phi_m \ , \quad 
F_{mn}=i\lambda^{-1}[\Phi_m,\Phi_n] \ 
\nonumber \\
\end{eqnarray}
and consequently
\begin{eqnarray}\label{dethelp1}
& &\det\left(\begin{array}{cc}
	0 & \tau_{
\hnu}+\tau_{\ty}D_{\hnu}\Phi_{\ty} \\
	\bar{\tau}_{\hmu}+D_{\hmu} \ty \bar{\tau}_{\ty} & 
	H_{\hmu\hnu}+\mB_{\hmu\hnu}+\lambda F_{\hmu\hnu} \end{array}\right)=
	\nonumber \\
& & =\det\left(\begin{array}{ccc}
0 &\tau_{\bnu}+\tau_{\ty}D_{\bnu}\Phi_{\ty}& 
\tau_n+\tau_{\ty}i\lambda^{-1}[\Phi_{n},\Phi_{\ty}] \\
\bar{\tau}_{\bmu}+D_{\bmu}\Phi_{\ty}\bar{\tau}_{\ty} & 
H_{\bmu\bnu}+\mB_{\bmu\bnu}+\lambda F_{\bmu\bnu} &
H_{\bmu n}+\mB_{\bmu n}+D_{\bmu} \Phi_n \\
\bar{\tau}_m+i\lambda^{-1}[\Phi_m,\Phi_{\ty}]\bar{\tau}_{\ty} & 
H_{m\bnu}+\mB_{m\bnu}-D_{\bnu}\Phi_m & H_{mn}+\mB_{mn}+i
\lambda^{-1}[\Phi_m,\Phi_n] \\ \end{array}\right) \ .
\nonumber \\
	\end{eqnarray}
For simplicity we introduce notation $E_{\mu\nu}=H_{\mu\nu}+\mB_{\mu\nu}$. Further, let us now presume that the matrix $E_{mn}$	has an 
inverse matrix $\tE^{mn}$. This is certainly always possible to define, for example
in the simplest case when $k=1$ so that $H_{zz}$ is one-dimensional. 
Then we can write the determinant (\ref{dethelp1}) as
\begin{equation}
\det \left(\begin{array}{cc} 
\bA_{\ty\ty} & \bA_{\ty \bnu} \\
\bA_{\bmu \ty} &  \bA_{\bmu\bnu} \end{array}\right) \ , 
\end{equation}
where 
\begin{eqnarray}\label{bAcomponents}
& &\bA_{\ty\ty}=
-(\tau_m+i\lambda^{-1}[\Phi_m,\Phi_{\ty}]\tau_{\ty})
(Q^{-1})^{mn}
(\bar{\tau}_n+i\lambda^{-1}[\Phi_n,\Phi_{\ty}]\bar{\tau}_{\ty}) \ , \nonumber \\
& & \bA_{\ty\bnu}=\tau_{\bnu}+\tau_{\ty}D_{\bnu}\Phi_{\ty}
-(\tau_n+i\lambda^{-1}\tau_{\ty}[\Phi_n,\Phi_{\ty}])(Q^{-1})^{nm}
(E_{m\bnu}-D_{\bnu}\Phi_m) \  , \nonumber \\
& &\bA_{\bmu \ty}=\bar{\tau}_{\bmu}+D_{\bmu}\Phi_{\ty}\bar{\tau}_{\ty}
-(E_{\bmu m}+D_{\bmu}\Phi_m)(Q^{-1})^{mn}
(\bar{\tau}_n+i\lambda^{-1}[\Phi_n,\Phi_{\ty}]\bar{\tau}_{\ty}) \ , \nonumber \\
& &\bA_{\bmu\bnu}=E_{\bmu\bnu}-(E_{\bmu m}+D_{\bmu}\Phi_m)
(Q^{-1})^{mn}(E_{n\bnu}-D_{\bnu}\Phi_n) \ , \nonumber \\
\end{eqnarray}
where 
\begin{equation}
Q_{mn}=E_{mn}+i\lambda^{-1}[\Phi_m,\Phi_n]  \ , 
\end{equation}
and where $(Q^{-1})^{mn}$ is its inverse $Q_{mk}(Q^{-1})^{kn}=\delta_m^n$. 

We showed in previous section that T-dual of D(p-1)-brane is again non-relativistic
D(p-1-k)-brane on condition when components of $\tau$ along directions we dualize
are equal to zero. Let us then impose the same condition. Further, let us express final form of (\ref{bAcomponents}) using T-dual form of the metric given in 
(\ref{Tdualemtr}) and we obtain 
\begin{eqnarray}
& &\bA_{\ty\ty}=-\lambda^{-2}\tau_{\ty}[\Phi_{\ty},\Phi_m]
(Q^{-1})^{mn}[\Phi_n\Phi_{\ty}]\bar{\tau}_{\ty} \ , \nonumber \\
& & \bA_{\ty\bnu}=\tau_{\bnu}+\tau_{\ty}D_{\bnu}\Phi_{\ty}
-i\lambda^{-1}\tau_{\ty}[\Phi_{\ty},\Phi_n](Q^{-1})^n_{ \ m}(\tE^{m}_{ \ \bnu}
+\tE^{mk}D_{\bnu}\Phi_k)
 \  , \nonumber \\
& & \bA_{\bmu \ty}=
\bar{\tau}_{\bmu}+D_{\bmu}\Phi_{\ty}\bar{\tau}_{\ty}
-i\lambda^{-1}(\tE_{\bmu}^{ \  m}+D_{\bmu}\Phi_k\tE^{km})(Q^{-1})_m^{ \ n}
[\Phi_n,\Phi_{\ty}]\bar{\tau}_{\ty} \ , \nonumber \\
& & \bA_{\bmu\bnu}=P[\tE_{\bmu\bnu}] +P[\tE_{\bmu r}
\tE^{rs}((Q^{-1})_s^{\ t}-\delta_s^t)\tE_{t\bnu}]  \ , \nonumber \\
\end{eqnarray}
where  $P[\tE_{\bmu\bnu}]$ is pull-back of the T-dual metric defined as
\begin{equation}
P[\tE_{\bmu\bnu}]=\tE_{\bmu\bnu}+D_{\bmu}\Phi_m \tE^m_{ \ \bnu}+
\tE_{\bmu}^{ \ n}D_{\bnu}\Phi_n+D_{\bmu}\Phi_m \tE^{mn}
D_{\bnu}\Phi_m \ , 
\end{equation}
and where 
\begin{eqnarray}
& &P[\tE_{\bmu r}
\tE^{rs}((Q^{-1})_s^{\ t}-\delta_s^t)\tE_{t\bnu}]=\tE_{\bmu}^{ \ m}\tE_{nk}((Q^{-1})^k_r-\delta^k_r)\tE^r_{ \ \bnu}
+\tE_{\bmu}^{ \ n}\tE_{nl}
((Q^{-1})^l_k-\delta^l_k)\tE^{kn}D_{\bnu} \Phi_n\nonumber \\
& &+D_{\bmu}\Phi_n \tE^{nr}\tE_{rk}((Q^{-1})^k_s-\delta^k_s)\tE^s_{ \ \bnu}+
D_{\bmu} \Phi_k \tE^{kr}\tE_{rs}((Q^{-1})^s_t-\delta^s_t)\tE^{tn}D_{\bnu}\Phi_n) \ . \nonumber \\
\end{eqnarray}
Finally, $\tphi$ is given in (\ref{Tdualphi}).
Collecting all these terms together we obtain an action for $N$ non-relativistic D(8-k)-branes in the form 
%
%
\begin{equation}\label{SkNfinal}
S=-T_{8-k}\str\int d^{9-k}\xi e^{-\tphi}
\sqrt{-\det \left(\begin{array}{cc}
\bA_{\ty\ty} & \bA_{\ty \bnu} \\
\bA_{\bmu \ty} & \bA_{\bmu\bnu} \\ \end{array}\right)\det Q^l_k} \ . 
\end{equation}
This is the final for of the action for $N$ D(8-k)-branes. Observe that there is general non-zero $\bA_{\ty\ty}$ as opposite to the case of single D(8-k)-brane. 
However note that for collection $N$ D(8-k)-brane that are localized at single
point $\ty_0$ we have that $\Phi_{\ty}=\ty_0 \bI_{N\times N}$ where $\bI_{N\times N}$ is unit matrix. As a result $\Phi_{\ty}$ commutes with all matrices and hence  $\bA_{\ty\ty}=0$. Then the action  (\ref{SkNfinal}) has similar form as the action for single D(8-k)-brane in non-relativistic background which is nice consistency check.

\section{T-Duality Along $\ty-$Direction}\label{fifth}
Now we are ready to study how non-relativistic D(8-k)-brane transforms under
T-duality along directions transverse to its world-volume. Since T-duality along
spatial directions that are transverse to its world-volume is the same as in case
of relativistic Dp-brane we skip this analysis and recommend 
\cite{Kluson:2020ipe} for more details. Instead we focus on T-duality along
$\ty-$direction.

We begin with the simpler case which is  D8-brane
transverse to $\ty$-direction and perform T-duality along it. It is well known that 
in order to perform T-duality along transverse direction we should consider configuration of infinite number of D8-branes on covering space. To do this we should presume that $\ty$ coordinate is compact so that the covering space is real line. Then the the world-volume matrix valued fields obey  following quotient condition
\cite{Brace:1998ku}
\begin{eqnarray}\label{quotcon}
& &	\mU\Phi_{\ty} \mU^{-1}= \lambda^{1/2}+\Phi_{\ty} \ , 
	\nonumber \\
& &	\mU A_{\hmu}\mU^{-1}=A_{\hmu} \ . \nonumber \\
	\end{eqnarray}
	In order to solve quotient equation it is natural to introduce 
	an auxiliary Hilbert space of functions $f(y)$ on which $\Phi_{\ty}$ and $\mU$ act. 
	Then  $\mU$ is generator of the functions on this covering space with coordinate $y$ in the form  
	\begin{equation}
	\mU=e^{i\frac{y}{\sqrt{\lambda}}} \ , 
	\end{equation}
	where $y$ is coordinate of the space on which  functions 
	$f(y)$ are defined.  Then $\Phi_{\ty}$ has to be equal to
	\begin{equation}\label{Phiop}
	\Phi_{\ty}=i\lambda\partial_{y}-A_y(y) \ , 
	\end{equation}
	where now $A_y(y)$ is ordinary function  that acts on Hilbert space by 
	ordinary multiplication. 
Using (\ref{Phiop}) we obtain
\begin{equation}\label{quotcon1}
D_{\hmu}\Phi_{\ty}=-\lambda \mF_{\hmu y} \ , \quad \mF_{\hmu y}=\partial_{\hmu}A_y-
\partial_y A_{\hmu} \
\end{equation}
so that the action for $N$ D8-branes can be written as 
\begin{eqnarray}
S=-\frac{T_8}{\sqrt{\lambda}}\int d^9\xi dy
e^{-\hphi}
\sqrt{-\det \left(\begin{array}{cc}
	0 & \tau_{\hnu}+\lambda\tau_{\ty}\mF_{ y \hnu} \\
\bar{\tau}_{\hmu}-\lambda \mF_{\hmu y}\bar{\tau}_{\ty} & 
H_{\hmu\hnu}+\mB_{\hmu\hnu}+\lambda \mF_{\hmu\hnu} \\
\end{array}\right)}=\nonumber \\
=-T_9\int d^9\xi dy 
e^{-\hphi}\sqrt{-\bar{\tau}_{\ty}\tau_{\ty}}
\sqrt{-\det \left(\begin{array}{cc}
	0 & \tau_{\hnu}\tau_{\ty}^{-1}+\lambda \mF_{y\bnu} \\
-\bar{\tau}_{\hmu}\bar{\tau}_{\ty}^{-1}+\lambda \mF_{\hmu y} & 
E_{\hmu\hnu}+\lambda \mF_{\hmu\hnu} \\ \end{array}\right)}
\end{eqnarray}	
using also the fact that
\begin{equation}
\tr=\frac{1}{\sqrt{\lambda}}\int dy \ . 
\end{equation}
Remember definition of $\tau_{\ty},\bar{\tau}_{\ty}$ as
was given in section (\ref{second}) we obtain the action in the form
\begin{equation}
S=-T_9\int d^{10}\xi e^{-\phi}\sqrt{-\det 
\left(\begin{array}{cc}
0 & E_{y\hnu}+\lambda \mF_{y\hnu} \\
E_{\hmu y}+\lambda \mF_{\hmu y} & E_{\hmu\hnu}+
\lambda \mF_{\hmu\hnu} \\ \end{array}\right)}
\end{equation}
which is the original form of DBI action for  D9-brane in the  the relativistic background with light-like isometry.
\subsection{T-duality in Case of D(8-k)-brane}
Now we proceed to the most interesting problem which is T-duality along 
$\ty$ direction that is transverse to the world-volume of non-relativistic
D(8-k)-brane. As is previous section we consider an array of infinite
number of  D(8-k)-branes whose world-volume fields obey quotient conditions
(\ref{quotcon}) and (\ref{quotcon1}). Note that there is additional condition 
on the matrix $\Phi_m$ in the form
\begin{eqnarray}
\mU \Phi_m \mU^{-1}=\Phi_m \ . \nonumber \\
\end{eqnarray}
Now $\Phi_m(\xi)$ and $A_{\bmu}(\xi)$ should be considered as ordinary functions $\phi_m(\xi,y)$ and $A_{\bmu}(\xi,y)$ defined on the space labelled 
by $y$  that act on test function $f(y)$ by ordinary multiplications. 

Then we again find that
\begin{equation}
D_{\bmu}\Phi_{\ty}=-\lambda \mF_{\bmu y} \ 
\end{equation}
together with 
\begin{equation}
i\lambda^{-1}[\Phi_{\ty},\Phi_m]=-\partial_y \phi_m
\end{equation}
and also $[\Phi_m,\Phi_n]=[\phi_m(y),\phi_n(y)]=0$ since this is commutator of ordinary functions. As a result we get that $Q_{mn}=E_{mn}$ and
$(Q^{-1})^{mn}=\tE^{mn}$. Using these results we obtain that components of the matrix $\bA$ have the form 
\begin{eqnarray}
& &\bA_{\ty\ty}=-\tau_{\ty}\bar{\tau}_{\ty}
\partial_y \phi_m \tE^{mn}\partial_y \phi_n \ , \nonumber \\
& &\bA_{\ty \bnu}=
\bar{\tau}_{\bnu}+\lambda \tau_{\ty}\mF_{y\bnu}
+\partial_y\phi_n (\tE^n_{ \ \bnu}+\tE^{mk}\partial_{\bnu}\phi_k)\tau_{\ty} \ , 
\nonumber \\
& &\bA_{\bmu \ty}=\bar{\tau}_{\bmu}-\lambda \mF_{\bmu y}\bar{\tau}_{\ty}-(\tE_{\bmu}^{ \ m}+\partial_{\bmu}\phi_k \tE^{km})\partial_y \phi_n\bar{\tau}_{\ty} \ , \nonumber \\
& &\bA_{\bmu\bnu}=P[\tE_{\bmu\bnu}] \ . \nonumber \\
\end{eqnarray}
Inserting these results into the action (\ref{SkNfinal}) we obtain that it has the form
\begin{eqnarray}
& &S=-\frac{T_{8-k}}{\lambda^{1/2}}
\int dy d^{9-k}\xi
e^{-\tphi}\sqrt{-\bar{\tau}_{\ty}\tau_{\ty}\det Q^i_j}\times \nonumber \\
& &\sqrt{-\det \left(\begin{array}{cc}
	\partial_y \phi_m\tE^{mn}\partial_y \phi_n &
	\tau_{\ty}^{-1}\bar{\tau}_{\bnu}+\partial_{y}\phi_n
	(\tE^n_{ \ \bnu}+\tE^{nk}\partial_{\bnu}\phi_k)+\lambda \mF_{y\bnu } \\
	-\bar{\tau}_{\ty}^{-1}\bar{\tau}_{\bmu}
	+(\tE_{\bmu}^{ \ m}+\partial_{\bmu}\phi_k\tE^{km})
	\partial_y \phi_m+\lambda \mF_{\bmu y} &  P[\tE_{\bmu\bnu}]+\lambda \mF_{\bmu\bnu}\\
	\end{array}\right)} \ . \nonumber \\
\end{eqnarray}
Finally using  $\tau_{\ty}=C, \bar{\tau}_{\ty}=-C$
we get that the action takes standard form of the relativistic D(9-k)-brane
action in the background with the light-like isometry which is nice consistency
check. Note also that the background fields are given by T-duality rules 
(\ref{Tdualemtr}).
\\
\\
{\bf Acknowledgement:}
\\
This work 
is supported by the grant “Integrable Deformations”
(GA20-04800S) from the Czech Science Foundation
(GACR).

\newpage

\end{document}